\documentclass{article}

\usepackage{PRIMEarxiv}

\usepackage[utf8]{inputenc} 
\usepackage[T1]{fontenc}    
\usepackage{hyperref}       
\usepackage{url}            
\usepackage{booktabs}       
\usepackage{amsfonts}       
\usepackage{nicefrac}       
\usepackage{microtype}      
\usepackage{lipsum}
\usepackage{fancyhdr}       
\usepackage{graphicx}       
\graphicspath{{media/}}     

\usepackage{dirtytalk}
\usepackage{booktabs}
\usepackage{listings}
\usepackage[normalem]{ulem}
\usepackage{enumitem}

\title{Part-time Power Measurements: nvidia-smi's Lack of Attention
}

\author{
  Zeyu Yang \\
  Department of Engineering Science \\
  University of Oxford \\
  Oxford, UK \\
  \texttt{zeyu.yang@eng.ox.ac.uk} \\
  \And
  Karel Ad\'{a}mek \\
  Department of Engineering Science \\
  University of Oxford \\
  Oxford, UK \\
  \texttt{karel.adamek@eng.ox.ac.uk} \\
  \And
  Wesley Armour \\
  Department of Engineering Science \\
  University of Oxford \\
  Oxford, UK \\
  \texttt{wes.armour@oerc.ox.ac.uk} \\
}

\begin{document}
\maketitle

\begin{abstract}
The Graphical processing unit (GPU), initially developed for graphics rendering, has emerged as the go-to accelerator for high throughput and parallel workloads, spanning scientific simulations to AI, thanks to its performance and power efficiency. Given that 6 out of the top 10 fastest supercomputers in the world use NVIDIA GPUs and many AI companies each employ 10,000’s of NVIDIA GPUs, an accurate understanding of GPU power consumption is essential for making progress to further improve its efficiency. Despite the limited documentation and the lack of understanding of its mechanisms, NVIDIA GPUs' built-in power sensor, providing easily accessible power readings via the nvidia-smi interface, is widely used in energy efficient computing research on GPUs. Our study seeks to elucidate the internal mechanisms of the power readings provided by nvidia-smi and assess the accuracy of the power and energy consumption data gathered from this method. We have developed a suite of micro-benchmarks to profile the behaviour of nvidia-smi power readings and have evaluated them on over 70 different GPUs from all architectural generations since power measurement was first introduced in the `Fermi' generation of GPU. We have identified several unique and unforeseen problems in terms of power/energy measurement using nvidia-smi, for example on the A100 and H100 GPUs only 25\% of the runtime is sampled for power consumption, during the other 75\% of the time, the GPU can be using drastically different power and nvidia-smi and results presented by it are unaware of this. This along with other findings can lead to a drastic under/overestimation of energy consumed, especially when considering data centres housing tens of thousands of GPUs. We proposed several good practices that help to mitigate these problems. By comparing our results to those measured from an external power-meter, we have reduced the error in the energy measurement by an average of 35\% and in some cases by as much as 65\% in the test cases we present. We have encapsulated our learning, our micro-benchmark and measurement good practice into a Python library for public use.
\end{abstract}


\section{Introduction}
\begin{figure}[h!]
    \centering
    \includegraphics[width=0.8\columnwidth]{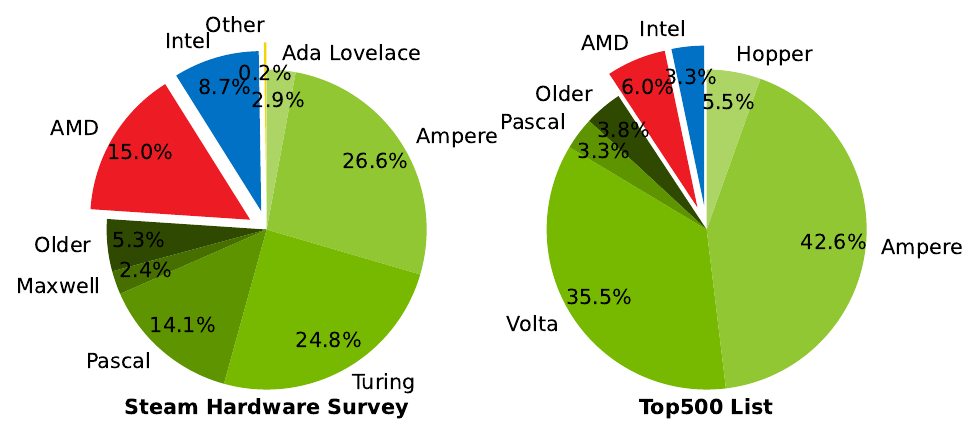}
    \caption{GPU market share among Steam users~\cite{steam_2023} and Top500 List~\cite{top500}. NVIDIA dominates both the Gaming and HPC market. Moreover, older architectures like the 'Turing', 'Volta', and 'Pascal' GPUs, launched between 2016 and 2018, still contribute a significant portion to the total market share.}
    \label{steam_share}
\end{figure}

GPUs are now critical components in computer systems, with NVIDIA as the leading player in the GPU market. Six out of the ten most powerful supercomputers in the TOP500 List~\cite{top500} use NVIDIA GPUs, together they consist of 82,240 GPUs. In fact 51 out of the first 100 supercomputers use NVIDIA GPUs, each featuring thousands to tens of thousand of GPUs. NVIDIA holds an 80.2\% market share in the discrete desktop GPU sector as of Q2 2023~\cite{Dow2023} and a commanding 91.4\% in the data center GPU market~\cite{Rau2022}. As of July 2023, on the video game distribution service Steam, 76.05\% of their users use NVIDIA GPUs~\cite{steam_2023}. Notably, older generation GPUs like `Turing', 'Volta' and `Pascal' still account for nearly half of the GPUs in both desktop and supercomputer environments, as Fig.~\ref{steam_share} illustrates. NVIDIA's market influence is further highlighted by its 1 trillion dollar market cap achieved in May 2023, with Q2 2023 revenues of \$13.5 billion~\cite{NVIDIA2023FR}. In this quarter, they have shipped over 5 million desktop GPUs~\cite{Dow2023} and more than 300,000 H100 GPUs~\cite{Shilov2023}.

\begin{figure}[t!]
    \centering
    \includegraphics[width=0.6\columnwidth]{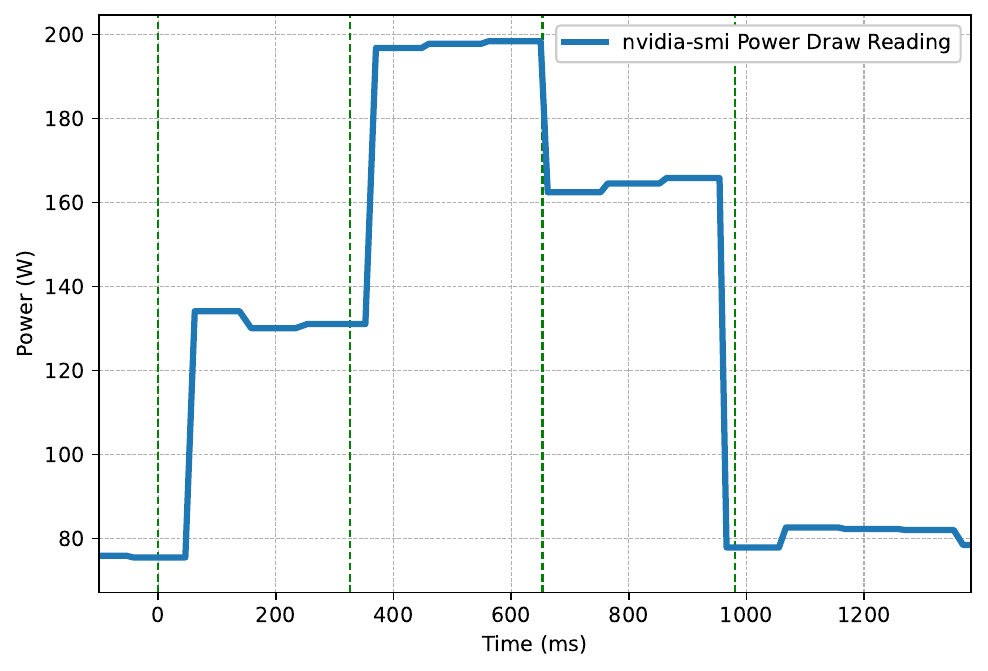}
    \caption{nvidia-smi can report drastically different power draw for the same CUDA Kernel, ranging from 80W to 200W. This is because nvidia-smi does not fully capture the power information on some GPUs. The figure shows a CUDA program that runs for 325ms on an A100 GPU. The kernel is executed 4 times, and the green dotted line indicates the beginning of each iteration.}
    \label{illustrate}
\end{figure}

Data centers were estimated to consume 1\% of global electricity~\cite{masanet2020recalibrating}, and contributed 0.3\% of emissions~\cite{jones2018stop} in 2018. Frontier, the first and only exascale supercomputer consumes 23 Megawatts of power. A significant workload in cloud GPUs is Machine Learning Training and Inference. A study~\cite{thompson2021deep} projected that training a top-tier ImageNet model could emit as much carbon as New York City does in a month, following current trends in accuracy, GPU-hours, and energy usage. Patterson et al.~\cite{patterson2022carbon} argued that this is a drastic overestimation, and suggested far lower carbon emissions. Furthermore, they also reported that Machine Learning workloads accounted for 10-15\% of Google's total energy usage from 2019 to 2021. While the proportion seems constant over time, Google's total energy consumption has been growing steadily by 20\% each year, from 12,750 GWh in 2019 to 22,289 GWh in 2022~\cite{google_env_report_2023}. With this trend, and AI/ML becoming ever more popular, energy consumption will continue to grow. Moreover, the computational power required for a widespread autonomous vehicle adoption would require as much electricity as all current data centres combined~\cite{sudhakar2022data}.

Despite the widespread usage of NVIDIA GPUs across these diverse fields, a glaring issue that emerges is the extensive power consumption. This highlights the critical need for energy efficiency. Adopting energy-efficient practices is essential both economically and environmentally."

Higher energy efficiency yields significant advantages across various domains. It typically results in lower overall energy use, reduced peak power draw, and decreased heat output. In data centers these reductions translate into lower operational costs, encompassing both power and cooling expenses, and consequently, a smaller carbon footprint. Simpler power and cooling solutions facilitate easier performance scaling, while lower operating temperatures boost device stability and longevity. For portable devices, improved energy efficiency can mean greater computational power and extended battery life in a more compact, lightweight design, improving user experience~\cite{yang2022instinctive}. It also opens possibilities for deployment in previously untenable scenarios due to power limitations.

Nonetheless, the prerequisite for enhancing energy efficiency is an accurate grasp of energy consumption. Research indicates that optimising algorithms based on inaccurate energy data can paradoxically lead to an 84\% increase in energy usage~\cite{fahad2019comparative}. Thus, to foster research and development in energy efficiency, an easily accessible, convenient, reliable, and precise method for measuring power is critically needed.

NVIDIA GPUs' onboard power sensors have the potential to fulfill these criteria. Yet, a significant gap in understanding their internal mechanisms hampers researchers' confidence in using data obtained from them. For example, as shown in Fig.~\ref{illustrate}, nvidia-smi reports drastically different power draw readings for running the same CUDA kernel.

This paper aims to investigate critical questions about the onboard sensors: What is being measured (over what time and which components)? How is it measured? How frequently are the measurements being taken? And most importantly, how accurate are these measurements?

Our contributions are summarised as follows:
\begin{itemize}
  \item Verified what is being measured and how is power measured electronically.
  \item Reverse engineered the internal data processing mechanisms within the nvidia-smi.
  \item Evaluated the Power and Energy measurement accuracy against those obtained from an external power meter.
  \item Proposed a set of measurement best practices capable of reducing energy measurement errors by up to 35\%.
\end{itemize}

Some of the key findings of our work includes:
\begin{itemize}
  \item The error in nvidia-smi's power draw is $\pm$5\% as opposed to $\pm$5W claimed by NVIDIA. On modern GPUs capable of drawing 700W this could lead to a $\pm$30W of over/underestimation. For a data centre with 10,000 GPUs, this would lead to an extra \$1 million in electricity cost yearly.
  \item On the A100 and H100 GPUs (and potentially future GPUs) only 25\% of the runtime is sampled for power consumption, during the other 75\% of the time, the GPU can be using drastically different power and nvidia-smi and results presented by it are unaware of this. 
\end{itemize}

We aim for our comprehensive analysis of NVIDIA GPU's onboard power sensors to empower researchers with accurate energy consumption data, leading to the creation of more energy-efficient algorithms and systems. Our investigation seeks to lay the groundwork for a standardised, reliable, and precise approach to power measurement, thereby advancing research and development in the vital area of energy efficiency.

\section{Background and Motivation}
In this section, we offer background information on electronically how is power being delivered to and measured on the GPU. We evaluate existing methods of power and energy measurement and review various studies that have utilised NVIDIA GPU's onboard power sensors.


\subsection{GPU Power Delivery}
PCIe form factor GPUs are powered via two sources. The majority of the power is supplied from the PC Power Supply Unit (PSU) via power cables at 12 volts. The most common are the 6 and 8-pin PCIe power cable, rated for 75W and 150W respectively. As modern GPUs become more power hungry, it is common to see cards with several of these connectors. The newest generation GPU uses the 12-pin PCIe gen 5 power connector, rated for 600W. Most of the data centre (Tesla) GPUs use the 384W rated 8-pin EPS connector instead. The other power source is the PCIe x16 slot itself, capable of supplying 75W, of which 10W is supplied via the 3.3V rail.

\subsection{Ways to measure power}
\paragraph{External Power Meters} Power is the product of Voltage and Current. While Voltage can be measured by an Analog-to-Digital converter (ADC), Current measurement needs additional steps to convert current to a Voltage representation. One common method involves measuring the magnetic field induced by the current using Hall Effect magnetic sensors, and then measure the voltage output of the sensor. Arafa et al.~\cite{arafa2020verified} used a Clamp Meter to measure the instruction level energy consumption of NVIDIA GPUs. While non-invasive, clamp meters are considered to be less accurate, and more suitable for high voltage applications. The other method is to pass the current through a shunt resistor with low resistance and measure the voltage across it (Ohm's law). This method is simple and robust, and widely used in low voltage applications.

Using the Shunt Resistor method, researchers have developed several power meters over the years~\cite{bedard2010powermon,laros2013powerinsight,romein2018powersensor}. NVIDIA themselves have a product called the ``Power Capture Analysis Tool (PCAT)'', as a part of their reviewer toolkit~\cite{pcat}, but it's not available for sale and primarily distributed to product reviewers like YouTubers. We purchased a similar product by ElmorLabs called ``Power Measurement Device (PMD)''~\cite{pmd} to serve the purpose.

\paragraph{Predictive Models}
Researchers have developed power models that predict the energy consumption of GPUs using a combination of architectural parameters, code/instruction analysis and performance counters. Leng et al.~\cite{leng2013gpuwattch} proposed a configurable, cycle-level capable model based on a bottom-up methodology from micro-architectural component parameters. Kandiah et al.~\cite{kandiah2021accelwattch} expanded upon the work to account for modern GPU architectures and Dynamic Voltage and Frequency Scaling (DVFS). Researchers also tried to use Machine Learning to model the GPU power consumption. Nagasaka et al.~\cite{nagasaka2010statistical} trained a linear regression model with performance counters as the independent variables, and power consumption as the dependent variable. Barun et al.~\cite{braun2020simple} built theirs based on random forests from 189 different kernels.

\begin{figure}[h!]
    \centering
    \includegraphics[width=0.6\columnwidth]{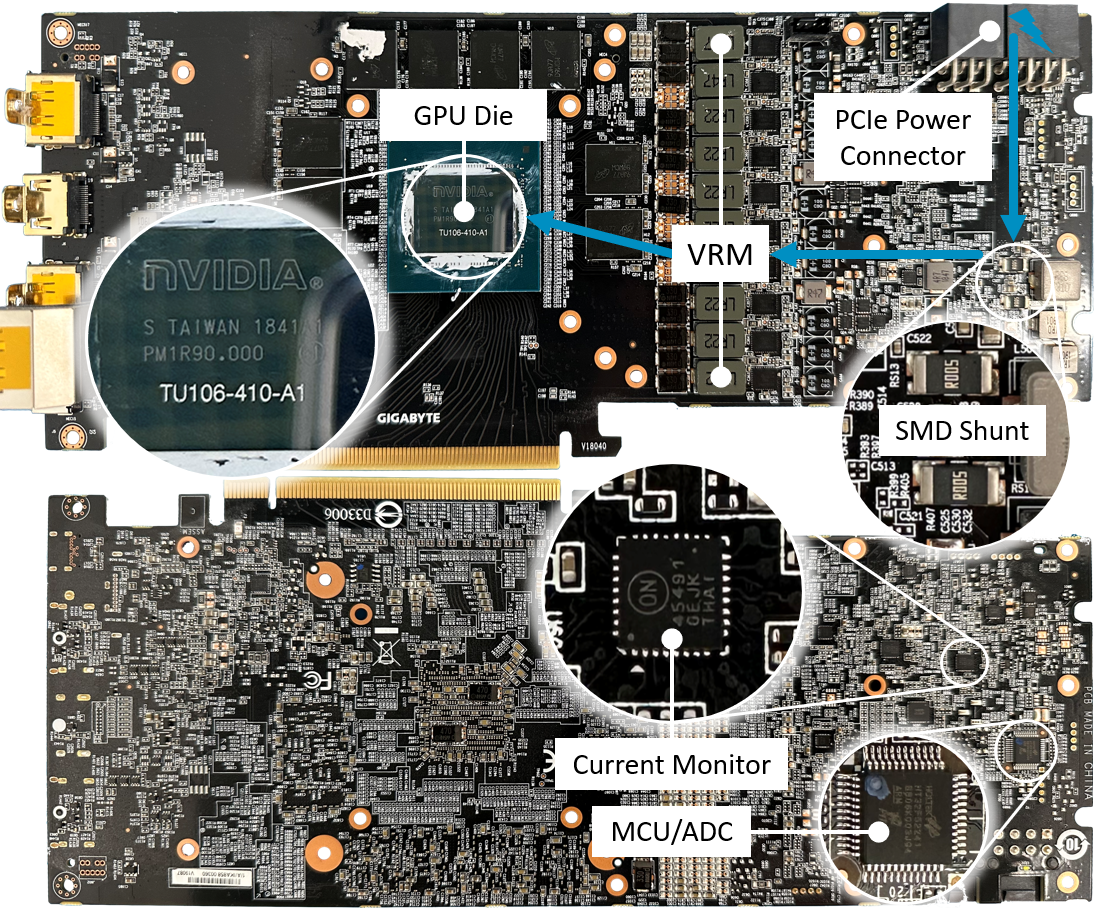}
    \caption{Front (above) and back (below) image of the PCB of the Gefore RTX 2060 Super GPU (by Gigabyte), as shown by the chip model TU106 on the GPU die. The 5m$\Omega$ SMD shunt resistors are located on the front, whereas the ON Semiconductor NCP45491 Current Monitor IC and the HOLTEK HT32F52241 32-Bit Arm Cortex-M0+ MCU is located at the back. The blue colored arrows illustrate the flow of eccentricity from the 6/8-pin PCIe power connector, through the shunt resistors, to the VRMs and reaches the GPU die.}
    \label{gpu_pcb}
    \vspace{-0.4cm}
\end{figure}

\subsection{GPU on-board power measurement}
NVIDIA GPUs have evolved from estimation-based to measurement-based on-board power tracking. As detailed in an NVIDIA patent~\cite{cha2011power}, older and less expensive models estimate power consumption by monitoring activity signals, linked to the number of active flip-flops, from different GPU functional blocks. Newer and higher-end models employ a measurement-based approach using the shunt resistor method. Extreme overclocking enthusiasts have exploited this by shorting the shunt resistors—known as 'shunt mod'~\cite{shuntmod}—to increase power draw beyond the GPU's TDP for liquid nitrogen overclocking.

We disassembled several GPUs to examine the PCB. We also examined high-res PCB images via web searches. We observed the existence of shunt resistors and current monitor ICs on GPUs from `Kepler' onward. For GPUs from the `Fermi' architecture and some other less expansive cards (e.g. Quadro K620) we found no evidence of those components. Fig.~\ref{gpu_pcb} shows the PCB of a RTX 2060S GPU that we disassembled. The shunt resistor and current monitor IC are clearly visible.

\subsection{nvidia-smi}
nvidia-smi (NVIDIA System Management Interface) is a command-line utility included in the GPU device driver. It is used for monitoring, managing, and querying GPU statistics. Running the command with the `--help-query-gpu' flag will show all the querying options available.

For power consumption data, on drivers released before March 30, 2023, there was only one option:

\begin{enumerate}
    \item \say{\textless{}power.draw\textgreater{} The last measured power draw for the entire board, in watts. This reading is accurate to within +/- 5 watts.}
\end{enumerate}

However, on drivers that were released after that, two new options were added. The documentation is also modified, adding more information, but making it more complicated to understand. The 3 options now are:

\begin{enumerate}
    \item \say{\textless{}power.draw\textgreater{} The last measured power draw for the entire board, in watts. On Ampere or newer devices, returns average power draw over 1 sec. On older devices, returns instantaneous power draw. This reading is accurate to within +/- 5 watts.}
    \item \say{\textless{}power.draw.average\textgreater{} The last measured average power draw for the entire board, in watts. Only available on Ampere (except GA100) or newer devices. This reading is accurate to within +/- 5 watts.}
    \item \say{\textless{}power.draw.instant\textgreater{} The last measured instant power draw for the entire board, in watts. This reading is accurate to within +/- 5 watts.}
\end{enumerate}

We will investigate and comment on each of the claims made in NVIDIA's documentation in the investigation chapter.

While the NVIDIA Management Library (NVML)~\cite{nvml} also provides power readings, at the time of writing, the documentation for NVML is outdated, lacking details on newly added power draw options. Conversely, nvidia-smi, which utilises NVML, reports similar power readings. Our tests show that nvidia-smi's results align with other NVML-based research. Moreover, as a standard command line tool included with the driver and featuring auto logging, nvidia-smi offers a more convenient and standard option than NVML.

\subsection{Advantages/Disadv. of on-board power measurement}
The on-board sensors come built into PCB, eliminating extra cost and installation complexities. Power readings are easily accessible via nvidia-smi or NVML, which comes with the driver, bypassing the need for sudo privileges, and complex data conversions, as is necessary for measuring CPUs via Intel RAPL. The hardware and software is standard across NVIDIA product lines, ensuring comparability of data between GPUs.

However, NVIDIA's closed-source nature raises concerns about the transparency of measurements and methods. Coupled with vague and outdated documentation, this casts doubt on the reliability of nvidia-smi's data. Our study aims to demystify these internal mechanisms and validate nvidia-smi's accuracy, contributing to research in energy-efficient computing.

\subsection{Work that used nvidia-smi/NVML to measure power/energy}
Thanks to the convenience, researchers have employed power data from nvidia-smi/NVML to develop GPU modelling tools and enhance the energy efficiency of specific algorithms or applications: Kandiah et al.~\cite{kandiah2021accelwattch} proposed a GPU power model for energy consumption simulation and prediction. Arunkumar et al.~\cite{arunkumar2019understanding} investigated the energy efficiency for future multi-module GPUs; Adamek et al.~\cite{adamek2021efficiency} reduced the energy consumption of the Fast Fourier Transform (FFT) algorithm on GPUs using Dynamic Voltage and Frequency Scaling (DVFS); White et al.~\cite{white2022cutting} optimized the energy efficiency of Radio Astronomy signal processing by mixed-precision computing; Henderson et al.~\cite{henderson2020towards} studied energy consumption and carbon emission of Machine Learning; Jahanshahi et al.~\cite{jahanshahi2020gpu} characterised energy efficiency of multi-GPU ML inference servers, and Yu et al.~\cite{yu2023know} proposed a resource management strategy to save the energy consumption of cloud scale inference services.

In the seven studies reviewed, five merely mentioned using nvidia-smi or NVML for power/energy measurements without detailed methodology. The other two offered some specifics but assumed nvidia-smi as a uniform sampler. Any error margin cited quoted the 5W from NVIDIA's documentation. This situation highlights that researchers may have placed excessive trust in nvidia-smi, overlooking the critical importance of measurement methodology in their analyses.

\subsection{Different measurement needs}
We classify the measurement needs into two categories: Power and Energy. Average and maximum power consumption is needed when designing the power delivery and cooling for a system, whether it be a data center or a embedded device. Energy consumption over a period of time is needed for people interested in cost of the computation, for example the operating cost of a data center and the battery life of a mobile device. We will evaluate NVIDIA GPU's on-board power sensor and nvidia-smi's ability in both measurement scenarios.

\section{Equipment and Experimental Setup}
In this section we provide and justify all the equipment involved in our investigation, and introduce the custom benchmark load that we designed.

\subsection{GPU Selection}
All other related works individually tested a limited number of GPUs~\cite{burtscher2014measuring,aslan2022study,fahad2019comparative,sen2018quality,jay2023experimental}, making it difficult to generalise their results for all GPU models. While it's impossible to test all of the different GPU models (over a thousand and counting), our aim was to include GPUs from each different architecture generation, product line-ups, form-factors and manufacturers.

We tested over 25 different GPU models, and over 70 different GPUs. These GPUs span across all 12 architectural generations that support power management from `Fermi' to `Hopper', all 3 product lines (Tesla, Quadro, and GeForce), and different form factors (PCIe, SXM and Mobile). The selection includes flagship products such as the RTX 4090, RTX 3090, H100, and A100. Multiple different cards and form-factors were tested for key GPU models to ensure no outliers: 5 RTX 3090s, 10 H100s and 10 A100s were tested. Furthermore, among the 10 A100s, 2 of them were the SXM4-40GB variant, 4 were PCIe-80GB, and 4 were PCIe-40GB; among the 5 RTX 3090s, 4 were from Dell and 1 was from EVGA. n addition, we investigated the Grace Hopper Superchip, a hybrid CPU+GPU in the same package. A full list of GPUs tested is shown in Table.~\ref{table:gpus}.

\vspace{-0.3cm}
\begin{table}[htbp!]
\caption{List of GPU tested}
\centering
\label{table:gpus}
\begin{tabular}{l l c | l c | l c }
\toprule
\textbf{Architecture} & \textbf{Tesla} & \textbf{\#} & \textbf{Quadro} & \textbf{\#} & \textbf{GeForce} & \textbf{\#} \\ 
        & (Data Center) & & (Pro W/S)  & & (Gaming) & \\
\midrule
Hopper  & H100 & 10 & & & &   \\
        & GH200 480GB & 1 & & & &   \\
\midrule
Ada     &  &  &  &  & RTX 4090 & 1  \\
\midrule
Ampere  & A100 PCIE-40G & 4 & RTX A6000 & 10 & RTX 3090   & 5 \\
        & A100 PCIE-80G & 4 & RTX A5000 & 1 & RTX 3070 Ti & 1 \\
        & A100 SXM4-40G & 2 &           &   &            &   \\
        & A10           & 1 &           &   &             &  \\
\midrule
Turing  &               &   & RTX 8000  & 4 & TITAN RTX   & 4 \\
        &               &   &           &   & RTX 2080 Ti & 1 \\
        &               &   &           &   & RTX 2060 S  & 1  \\
        &               &   &           &   & GTX 1650 Ti & 1  \\
\midrule
Volta   & V100 SXM2-16G & 4 &           &   &             &  \\
        & V100 PCIE-16G & 1 &           &   &             &  \\
\midrule
Pascal  & P100 PCIE-16G & 5 &           &   & TITAN Xp & 1 \\
        &               &   &           &   & GTX 1080ti  & 1 \\
        &               &   &           &   & GTX 1080    & 1  \\
\midrule
Maxwell 2.0  & M40 & 1 &               &   & TITAN X & 1 \\
Maxwell 1.0  &         &   & K620      & 1 & GTX 745 & 1 \\
\midrule
Kepler 2.0   & K80     & 1 &           &   &    &   \\ 
Kepler 1.0   & K40     & 1 &           &   &    &   \\    
\midrule
Fermi 2.0   & M2090     & 1 &           &   &    &   \\ 
Fermi 1.0   & C2050     & 1 &           &   &    &   \\   
\bottomrule
\end{tabular}
\end{table}
\vspace{-0.4cm}

\begin{figure}[t!]
    \centering
    \includegraphics[width=0.6\columnwidth]{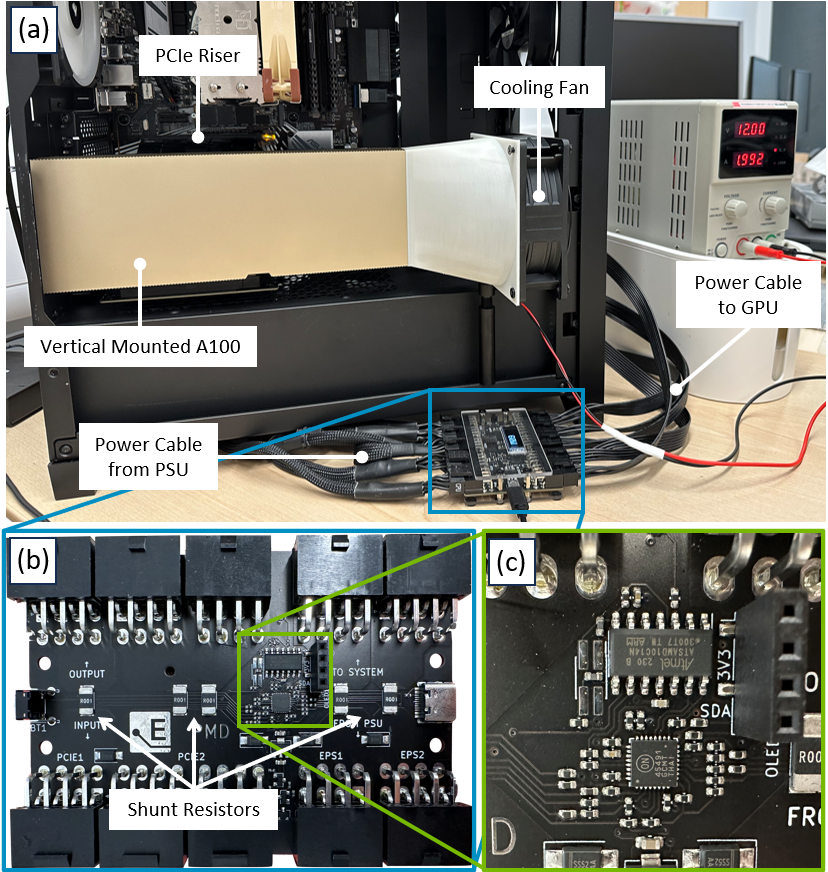}
    \caption{(a) The test-bench PC with the A100 GPU installed. (b) The PMD PCB, and showing the location of the 1m$\Omega$ shunt resistor. (c) Zoom in on the PMD PCB, showing the same ON Semiconductor NCP45491 used on many GPUs and a Atmel ARM MCU.}
    \label{setup}
    \vspace{-0.4cm}
\end{figure}

\subsection{The Power Measurement Device}
The Power Measurement Device, positioned between the power supply and GPU, channels all GPU currents through its current sensing shunt resistors. A separate PCIe riser is needed to measure power supplied via the PCIe x16 slot, which disconnects the power rails from the motherboard while allowing data flow, and replaces it with a 4-pin EPS connector. However, power from the 3.3V rail isn’t captured, potentially underestimating the GPU’s power draw by up to 10W.

The voltage and current(shunt voltage) are quantized by a 12-bit (4096-levels) ADC. The voltage range is 0-31V (0.007568V per level), and the current range is 0-200A (0.0488A per level). For a normal operating scenario of 12V and 12.5A, the resolution is 1586 levels for voltage and 256 levels for current. The voltage is rated to $\pm$0.1V and current is rated to $\pm$0.5A. The internal sampling frequency is 34 KHz, but it's limited by the serial data transmission speed from the PMD to a host PC. Consequently, the provided software updates at 10 Hz and is Windows-compatible only. We developed a data logger program to configure the PMD to send raw data at a baud rate of 921,600 to the host PC for later processing, achieving a sampling frequency of 5 KHz.

\subsection{Measurement Setup}
Fig.~\ref{setup}a shows the measurement setup for the A100 GPU as an example. The GPU needs to be vertically mounted with a PCIe extension cable to be able to install the PCIe riser PCB. Since the data centre cards are passively cooled, a 24W fan is used with a air duct to cool the GPU. 

The test-bench PC has a Intel Core i7-9700K CPU, 32G of dual channel DDR4 memory, a 850W PSU form Corsair, and runs Ubuntu 22.04 LTS Operating System. All the GPUs that we had physical access to were tested on this test-bench.

\subsection{Benchmark load}

\begin{figure}[h!]
    \centering
    \includegraphics[width=0.6\columnwidth]{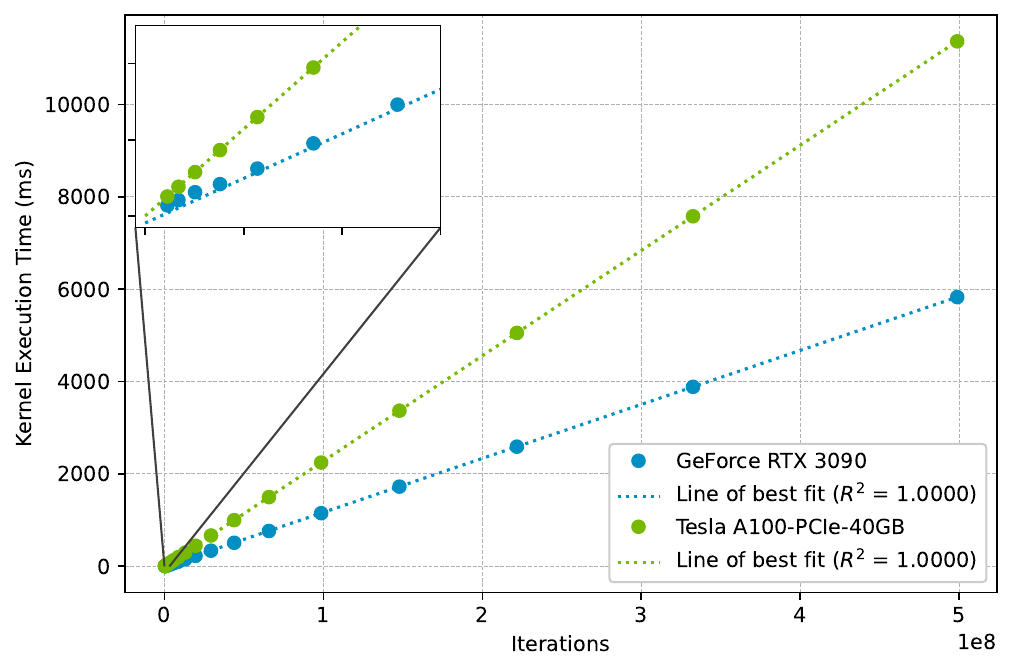}
    \caption{The number of iterations and kernel execution time demonstrates a linear relationship on both RTX 3090 and A100. Each marker represents a tested iteration count, and the dotted line illustrates the best fit line derived from linear regression. Both models exhibit an R-squared value of 1.000, indicating a perfect fit. The slope of the line is then used to control the duration of the high power state.}
    \label{scaling_params}
    \vspace{-0.4cm}
\end{figure}

The benchmark load is designed to induce high and low GPU power consumption states in a square wave pattern, and the amplitude (power draw), frequency and number of cycles can be precisely controlled. This serves as a stress test for probing the nvidia-smi's internal mechanisms. The low power state is achieved by a timed sleep, with the duration defining the state’s length, while the high power state is achieved through a custom CUDA kernel performing a data-dependent chain of vector Fused Multiply-Add (FMA) operations. The computation duration is linear to the length of the FMA chain (Fig.~\ref{scaling_params}). To control the duration of this high power state, linear regression was used to determine the gradient between the time measured for a set of arbitrary chain lengths. Amplitude control is achieved by activating different amounts of streaming multiprocessors (SM) of the GPU by setting the number of blocks to be a fraction of the total SM count of the GPU. Fig~\ref{ss_error_lin_regres} shows the different power draw levels can be precisely controlled. The code snipped is shown in Listing~\ref{load code}.

\begin{lstlisting}[language=C++, label=load code, caption={Benchmark load CUDA code. The CUDA kernel conducts a chain of vector multiplications and additions. Each operation depends on the data from the previous operation, ensuring sequential execution. The size of the vector is determined by the product of maximum threads per block and a streaming multiprocessor (SM) count. When executing the kernel, the number of threads (nthreads) is set to max threads per block, and number of blocks (nblocks) is set to the SM count.}]
__global__ void kernel(float *x, int niter) {
    int tid = threadIdx.x + blockDim.x * blockIdx.x;
    #pragma unroll
    for (int i=0; i<niter; i++) {
        x[tid] = x[tid] * 2 + 2;
        x[tid] = x[tid] / 2 - 1;
    }
}
int main(int argc, const char **argv) {
    int nblocks, nthreads, nsize;  float *d_x;
    cudaDeviceProp devProp = getDeviceProperties();
    nblocks = devProp.multiProcessorCount * PERCENT;
    nthreads = devProp.maxThreadsPerBlock;
    if (nblocks < 1) nblocks = 1;
    nsize = nblocks * nthreads;
    // cudaMalloc d_x array on GPU with nsize
    kernel<<<nblocks,nthreads>>>(d_x, DELAY*SCALE);
    cudaDeviceSynchronize();
    usleep(DELAY*1000);
}
\end{lstlisting}

\section{Investigation and Results}
We've conducted three experiments to investigate the internal mechanisms of nvidia-smi. We ran the experiments on the over 70 GPUs we've selected, and present comprehensive and exhaustive results.
\subsection{Sampling Frequency / Power update Frequency}
\begin{figure}[h!]
    \centering
    \includegraphics[width=0.65\columnwidth]{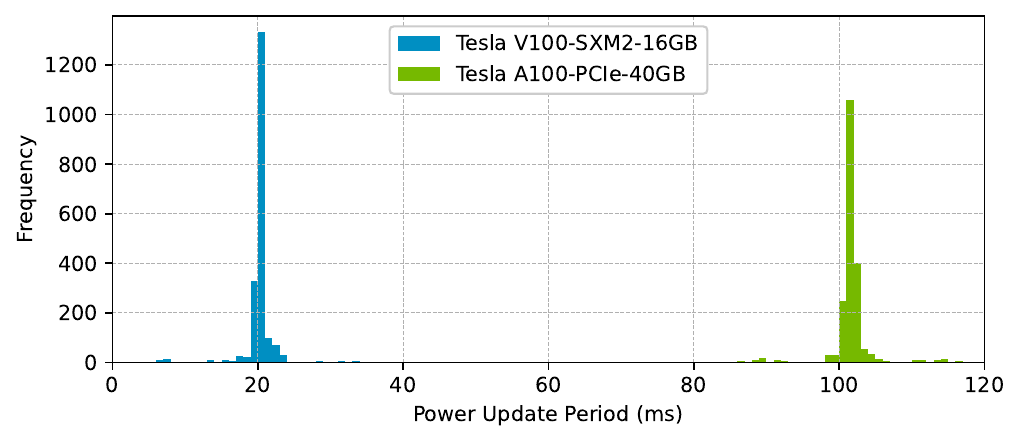}
    \caption{Histogram of the measured power update period for the V100 and A100 GPU. While there are fluctuations, majority of the update period is at 20ms for V100, and 101ms for A100.}
    \label{pwr_update_freq}
    \vspace{-0.4cm}
\end{figure}

When querying nvidia-smi, there is the option to specify a querying frequency/period in units of milliseconds, where the actual period can deviate by several milliseconds. However, it was observed that the power draw reading remains constant for several query samples before updating. We call this the Power Update Frequency/Period. To test this, we ran the benchmark load for a short period of time, with a 20ms square wave period to ensure variating activity on the GPU. Then, the length of each time period where the power draw reading remained the same were counted. We take the median value as the Power Update Period of the GPU. Fig.~\ref{pwr_update_freq} shows the Power Update Period of a tested V100 and A100 GPU.

If the reported value is simply an sample of the instantaneous power draw, this 10-50Hz sampling frequency is orders of magnitudes slower than the activities on the GPU, typically at a clock frequency of GHz. This indicates that majority of the information would not be captured. This issue is further investigated later in the chapter.

\subsection{Transient Response}
The transient response of the nvidia-smi's power measurement was then tested. The step function is generated by the benchmark load with a single period and a duration of 6 seconds. The rise time, defined as the time taken for the power draw to rise from 10\% to 90\% of max power draw, is measured. When the PMD is available (for the cards we have physical access to), the steady state error are also measured. 

The analysis of rise time and delay is crucial for quantifying nvidia-smi's power reading responsiveness to fluctuations in actual power consumption. The steady-state error analysis effectively decouples power readings from time-domain disturbances (e.g. delay and averaging), enabling accurate measurement of the inherent error in power draw values.

\begin{figure}[h!]
    \centering
    \includegraphics[width=0.75\columnwidth]{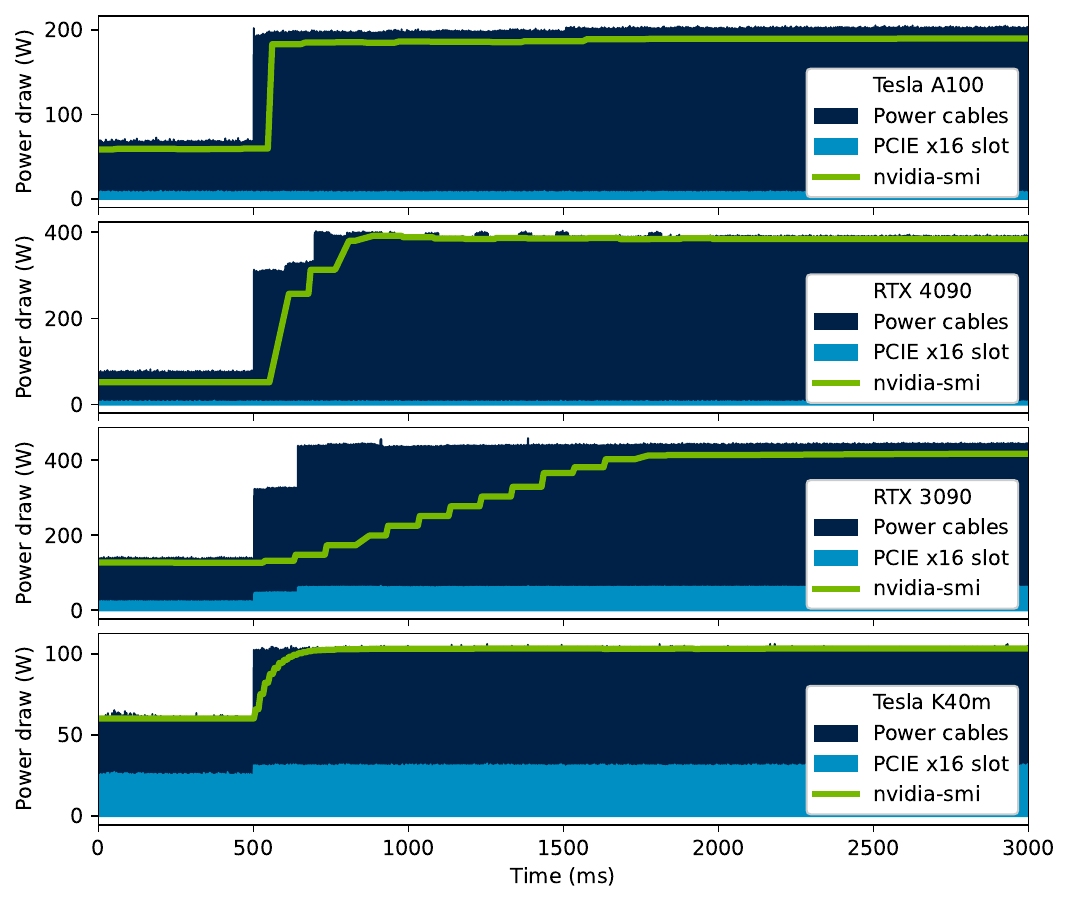}
    \caption{The 4 different kinds of transient response observed. The benchmark load began execution at 500ms.}
    \label{transient_response}
    \vspace{-0.4cm}
\end{figure}

\paragraph{Rise time and delay} Fig.~\ref{transient_response} shows the 4 different kinds transient response we observed. The first case is the actual power consumption rise time, it is nearly instant, and the nvidia-smi power consumption follows at the next power update, delayed anywhere between 0 to 100 milliseconds. The second case is that the actual power consumption takes several hundred milliseconds to rise, however the nvidia-smi power reading is still updated at the next power update interval. In contrast, in the third case, nvidia-smi clearly lags behind the actual power draw, with a linear growth over 1 second. The last case is a logarithmic growth over 200 milliseconds. 

The first 2 cases correspond to the \textless{}power.draw.instant\textgreater{} option in nvidia-smi, which represents \say{The last measured instant power draw}. Case 3 corresponds to the \textless{}power.draw.average\textgreater{} option, which is \say{The last measured average power draw over 1 sec}. Case 4, although lagging behind actual power draw, is the \textless{}power.draw.instant\textgreater{} option. This logarithmic growth is only observed in older GPUs of the `Kepler' and `Maxwell' generation.

These various responses imply that for some GPUs, if a short program is executed, the measured power are likely to be the activity happened before the program was executed.

\begin{figure}[h!]
    \centering
    \includegraphics[width=0.5\columnwidth]{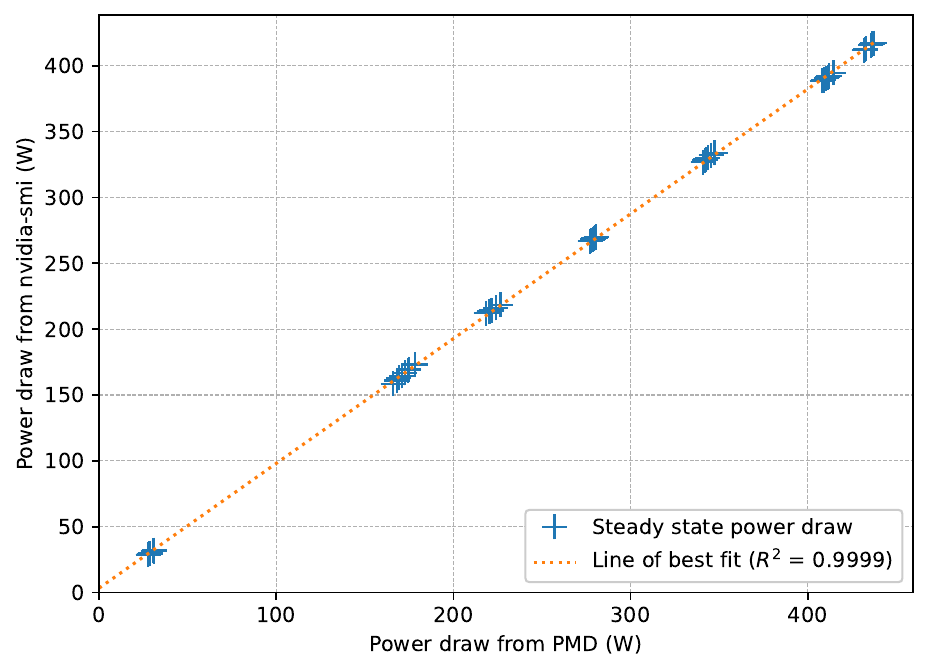}
    \caption{RTX 3090 steady state power draw from nvidia-smi plotted against PMD. 8 repetitions were tested for each of the 7 different power draw levels: Idle, 1\%, 20\%, 40\%, 60\%, 80\% and 100\% of total SM count, respectively corresponds to the clearly visible 7 clusters of points in the figure. The middle 5 clusters are roughly equally spaced apart. The Idle cluster on the left is further away since its on a lower GPU pstate, and the 100\% cluster on the right is less further apart due to the power limit at 420W. The dotted line is the line of best fit using linear regression with a R squared value of 0.9999.}
    \label{ss_error_lin_regres}
    \vspace{-0.4cm}
\end{figure}

\begin{figure}[t!]
    \centering
    \includegraphics[width=0.6\columnwidth]{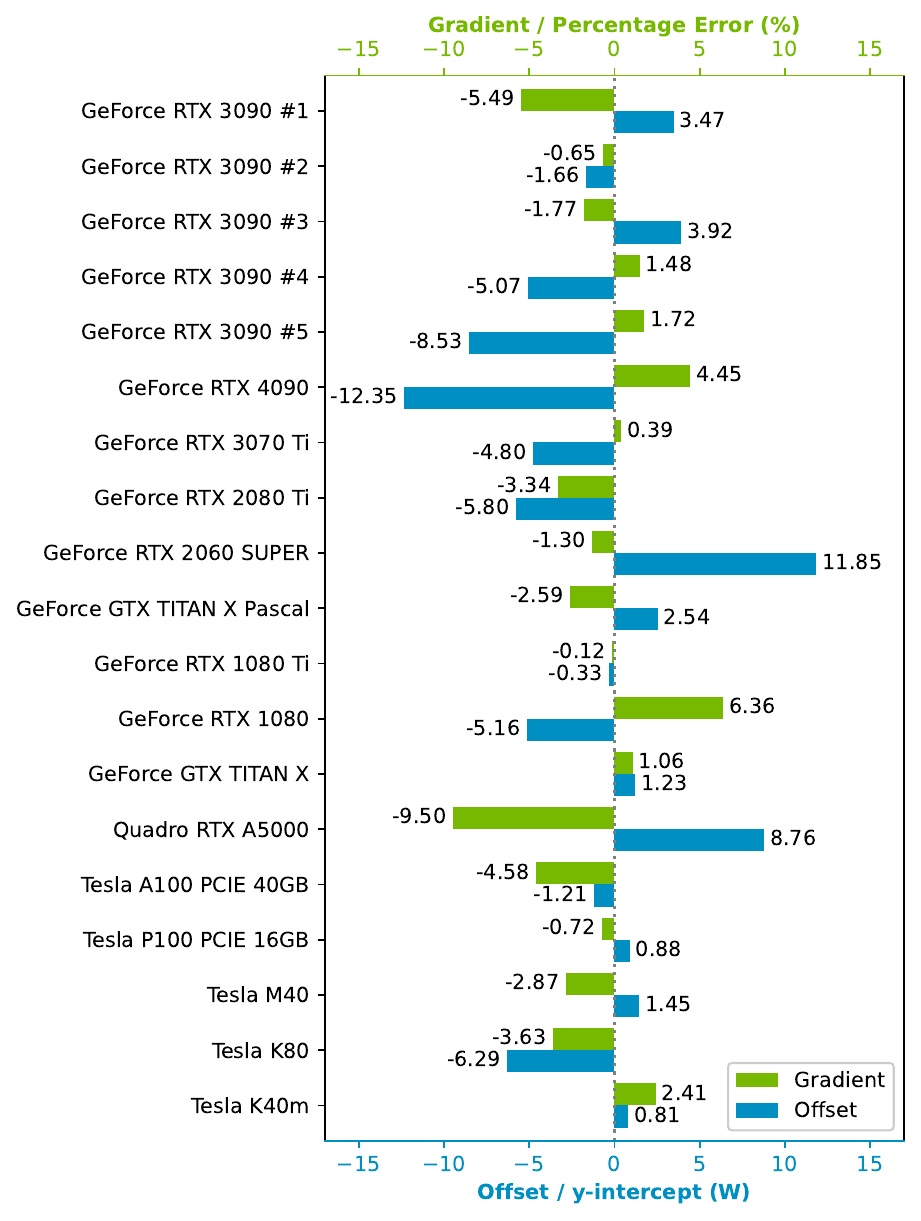}
    \caption{Gradient and offset of the steady state error of each GPU (with physical access) tested as shown in Fig.~\ref{ss_error_lin_regres}. The RTX 3090 \#1 is from EVGA, and \#2-5 are from Dell Alienware. RTX 2060 and 3070 are both from GIGABYTE. In some cases the gradient and offset are in the opposite direction, and can work against each other to reduce the overall error to some extent.}
    \label{ss_error}
    \vspace{-0.6cm}
\end{figure}

\paragraph{Steady State Error} The steady state error measures the error between the power draw reported by nvidia-smi and PMD when the GPU reaches a constant power draw. Different levels of power consumption were tested from idle to maximum. Fig.~\ref{ss_error_lin_regres} shows the result tested on a RTX 3090. There is a near perfect linear relationship between the two, however the gradient is not 1, indicating the error is proportional, rather than the flat $\pm$5W NVIDIA claimed. We repeated this experiment on all the GPUs we had physical access to, and plotted the gradient and y-intercept in Fig.~\ref{ss_error}. The tested GPUs, including identical/different models from the different/same manufacturers, showed no clear trends for any specific model or manufacturer.

The error is likely determined by a combination of manufacturer's component quality choice (magnitude of tolerance), and random error within the tolerance. In the majority of the cases the error is within $\pm$5\%, which disagrees with NVIDIA's claim that the power draw is with in $\pm$5W. A percentage error makes more sense since the shunt resistor that measures the current will have a resistance tolerance in percentage. Modern GPUs like the H100 has a TDP of 700W, under max load a 5\% error leads to a 35W error. If one trusted NVIDIA's documentation, 30W of power draw may be under/overestimated for a single GPU. For data centers with tens of thousands of GPUs, the magnitude of the error would be very significant.

Despite a deviation from a gradient of 1, the relationship remains almost perfectly linear for each GPU itself. This implies that when optimizing programs on the same GPUs, the direction of power efficiency improvement will be accurate, albeit with a minor magnitude error. It's improbable that nvidia-smi will misread an increase in power consumption as a decrease, thus it remains reliable for designing and optimizing power-efficient algorithms.

\subsection{Box-car Averaging window}
To determine whether the reported power draw is an instantaneous sample or an average over the past power update period, we can run multiple periods of the benchmark load with the square wave period matching the power update period. If the power draw reported is instantaneous, we should observe two distinct high and low power draws. Conversely, if the value is an average, the power draw should remain steady, halfway between the high and low states.

\begin{figure}[t!]
    \centering
    \includegraphics[width=0.8\columnwidth]{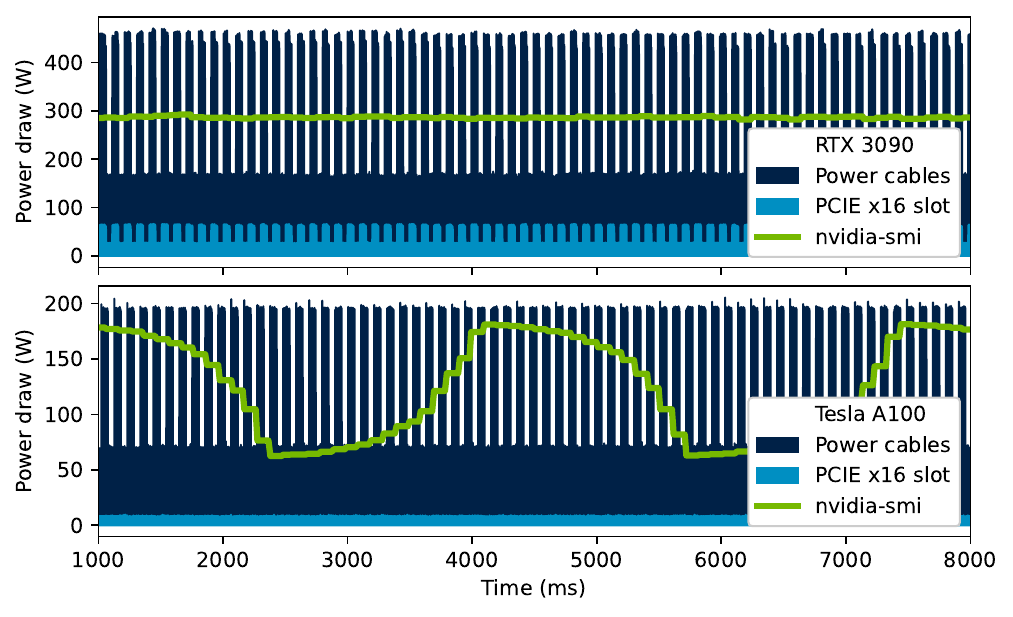}
    \caption{Running the benchmark load of a square wave period of 100ms on RTX 3090 and A100. The shaded area shows the actual power draw measured by PMD, whereas the green line shows the power draw reported by nvidia-smi. On RTX 3090 nvidia-smi's power stays flat in the middle, indicating the boxcar averaging window equals to the power update frequency. On A100 the nvidia-smi's power swings up and down, indicating the averaging window is a fraction of the power update period.}
    \label{5050}
    \vspace{-0.4cm}
\end{figure}

Fig.~\ref{5050} shows the results of this test on a RTX 3090 and a A100. On the RTX 3090, nvidia-smi’s power draw remained constant, indicating that the reported power draw value is an average of the past power update period. This is essentially a Boxcar averaging over time, whereas the size of the boxcar is the power update period. On the other hand, nvidia-smi’s power draw for the A100 fluctuated between high and low power consumption. Intermediate values between high and low power consumption exists indicates that the nvidia-smi's power is not an instantaneous sample, however the averaging boxcar window is not the entire power update frequency either. The only explanation would be the boxcar window is a fraction of the power update period. After some investigation, we found out that the period of the generated square wave load deviated slightly from exactly 100ms, and this created an aliasing effect between the actual power consumption and nvidia-smi’s update frequency, hence the fluctuation. Moreover, the starting point of the fluctuation is random among multiple runs because nvidia-smi starts measuring at boot time, and there is no way for the user to control the starting time of the averaging process nor synchronise with it.

To measure the boxcar averaging window, a indirect approach is employed. A model that emulates the boxcar averaging behaviour is created, taking nvidia-smi's power data and a window size as input. For each nvidia-smi sample timestamp, the average power of the specified window from the PMD data is calculated. The experiment is performed as follows:
\begin{enumerate}[leftmargin=*]
  \item Set benchmark load square wave period to a fraction of the power update period to create more aliasing effect. 
  \item Run benchmark load for 9 seconds and collect nvidia-smi and PMD power data.
  \item Reconstruct an emulated nvidia-smi data using the model.
  \item Discard the first second of data, and normalise both original and emulated nvidia-smi data to only compare the shape.
  \item Construct a loss function that calculates the MSE between the original and emulated nvidia-smi power data.
  \item Minimise the loss function using Nelder-Mead, with initial power window set as half of the power update frequency.
\end{enumerate}

\begin{figure}[t!]
    \centering
    \includegraphics[width=0.8\columnwidth]{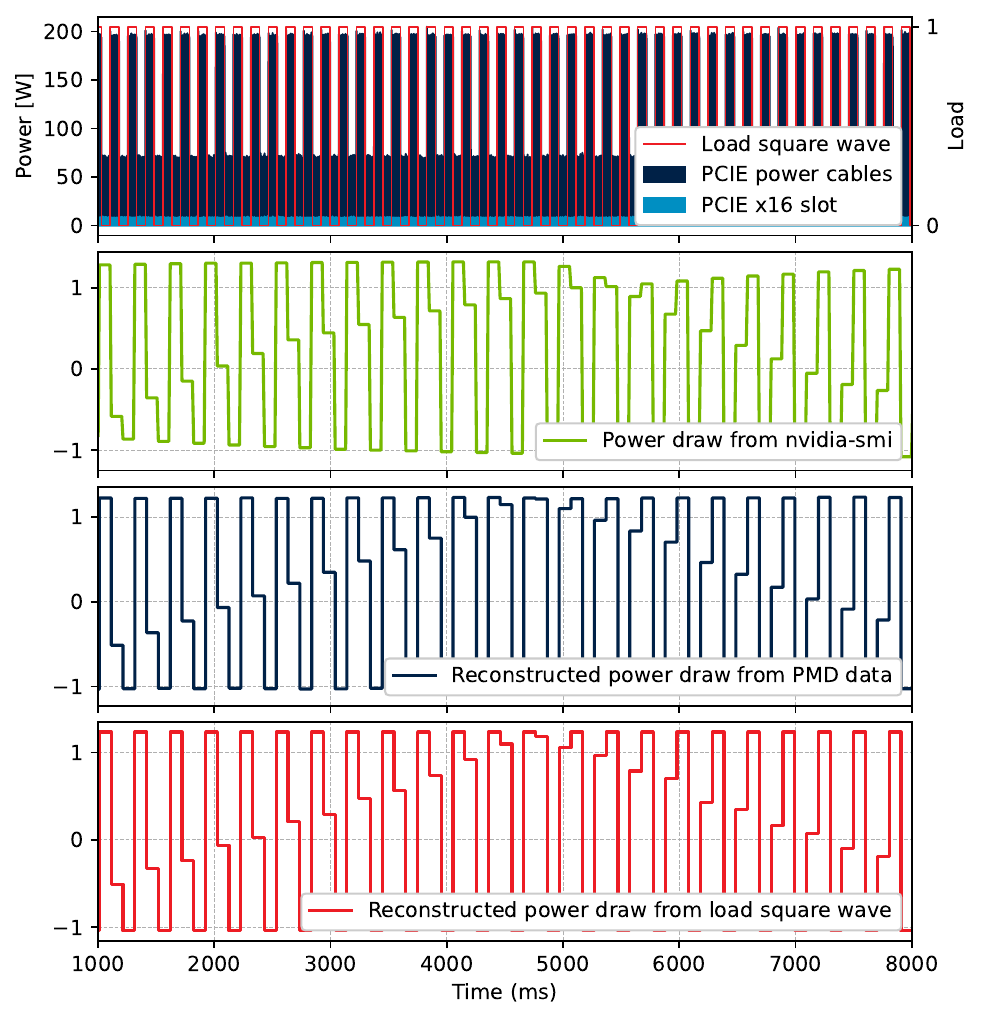}
    \caption{Running the benchmark load with period equal to 154ms on A100. Top plot shows the power draw data from PMD and the square wave load. The following 3 plots show the original nvidia-smi power data, emulated nvidia-smi data from PMD data, and emulated nvidia-smi data from square wave.}
    \label{A100_recontr}
\end{figure}

\begin{figure}[t!]
    \centering
    \includegraphics[width=0.65\columnwidth]{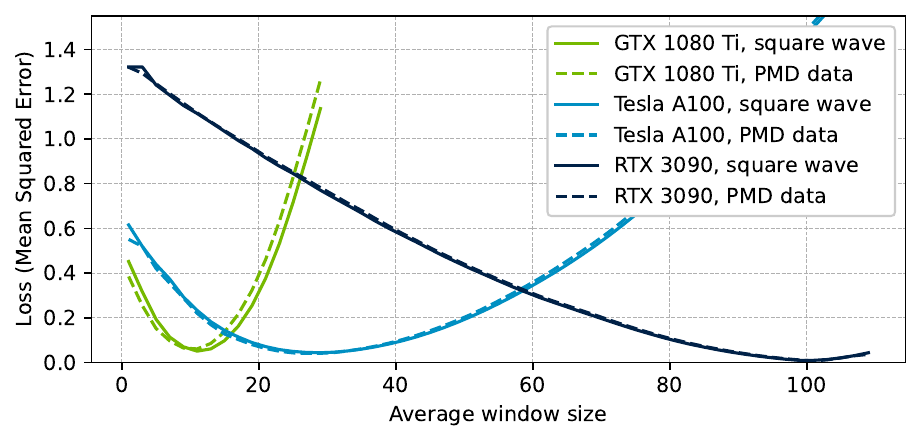}
    \caption{Loss function of 3 representative GPUs. Location of the minima is the same for emulation of nvidia-smi constructed by either PMD data or Benchmark Load square wave, indicating that the experiment can be performed on GPUs without PMD attached.}
    \label{loss_func}
    \vspace{-0.4cm}
\end{figure}

Fig.~\ref{A100_recontr} shows a particular run of the experiment on the A100. Since the actual power draw reflects the square wave load very well, using the square wave to reconstruct the nvidia-smi power data is nearly the same as using PMD data. This makes the experiment able to run on any GPU without the need of the PMD device. Fig~\ref{loss_func} shows the loss function of a particular run of 3 GPUs: GTX 1080 Ti, RTX 3090 and A100. The loss function suggests that the boxcar averaging window of GTX 1080 Ti is 10ms out of the 20ms power update frequency, A100 is 25/100ms, and RTX 3090 is 100/100ms.

\begin{figure}[t!]
    \centering
    \includegraphics[width=0.8\columnwidth]{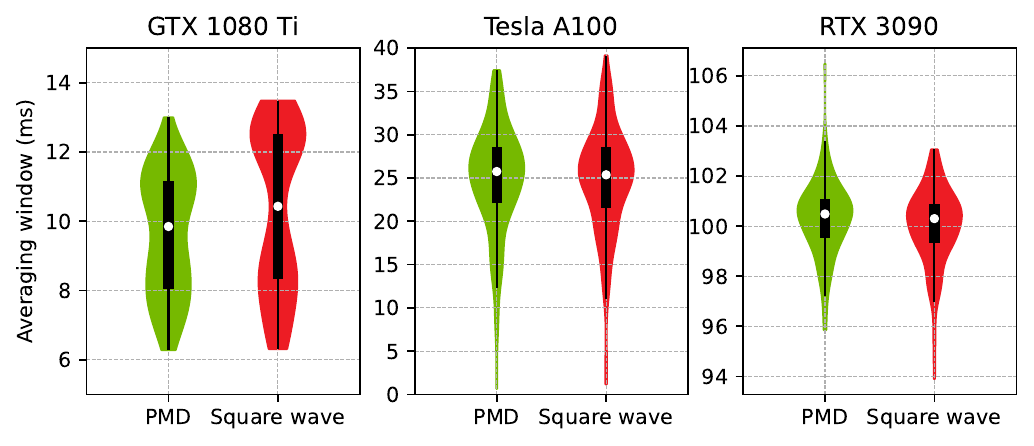}
    \caption{Violin plot of the distribution of the experiment results. The white dot shows the median, thick black bar shows the interquartile range (IQR), and the thin black line shows the upper and lower adjacent value. Standard Deviation of the PMD/Sqaure wave result for GTX 1080 Ti, A100 and RTX 3090 are 1.6/2.4, 3.3/3.2, and 1.2/1.3 respectively.}
    \label{violin}
    \vspace{-0.4cm}
\end{figure}

The above experiment is performed 32 times each for 6 different fractions: 2/3, 3/4, 4/5, 6/5, 5/4, and 4/3 of the power update period. Fig.~\ref{violin} shows the distribution of the results for the three GPUs. The noise in the data collected and outcome of the optimisation caused the results to deviate slightly.

\subsection{Overall Results}
The above 3 experiments: Power Update Frequency, Transient Response, and Averaging Window, were performed on all 70 GPUs we've tested. For the 20 GPUs we had physical access to, in addition to the steady state error analysis done using PMD data shown in Fig~\ref{ss_error}, we also investigated nvidia-smi's behaviour across different driver versions. Furthermore, to ensure there were no bias:
\begin{itemize}
    \item Multiple cards are tested for the same model.
    \item Cards from multiple manufactures were tested.
    \item Card in different form-factors were tested.
    \item Same card in different host machines were test.
\end{itemize}
The manufacturers of the GPU selection includes EVGA, PNY, GiGABYTE, Dell Alienware and Founder Edition cards. 5 RTX 3090s from 2 different manufacturers were tested in different host machines. 10 different H100 PCIE cards were tested, 8 from our university cluster and 2 from a cloud provider. Ten different A100 cards were tested, 2 are SXM4-40GB version from a cloud provider, 4 are PCIE-40GB from our university cluster, and 4 are immersion cooled PCIE-80GB cards. A GTX 1650Ti laptop is also tested.

\begin{figure}[t!]
    \centering
    \includegraphics[width=0.5\columnwidth]{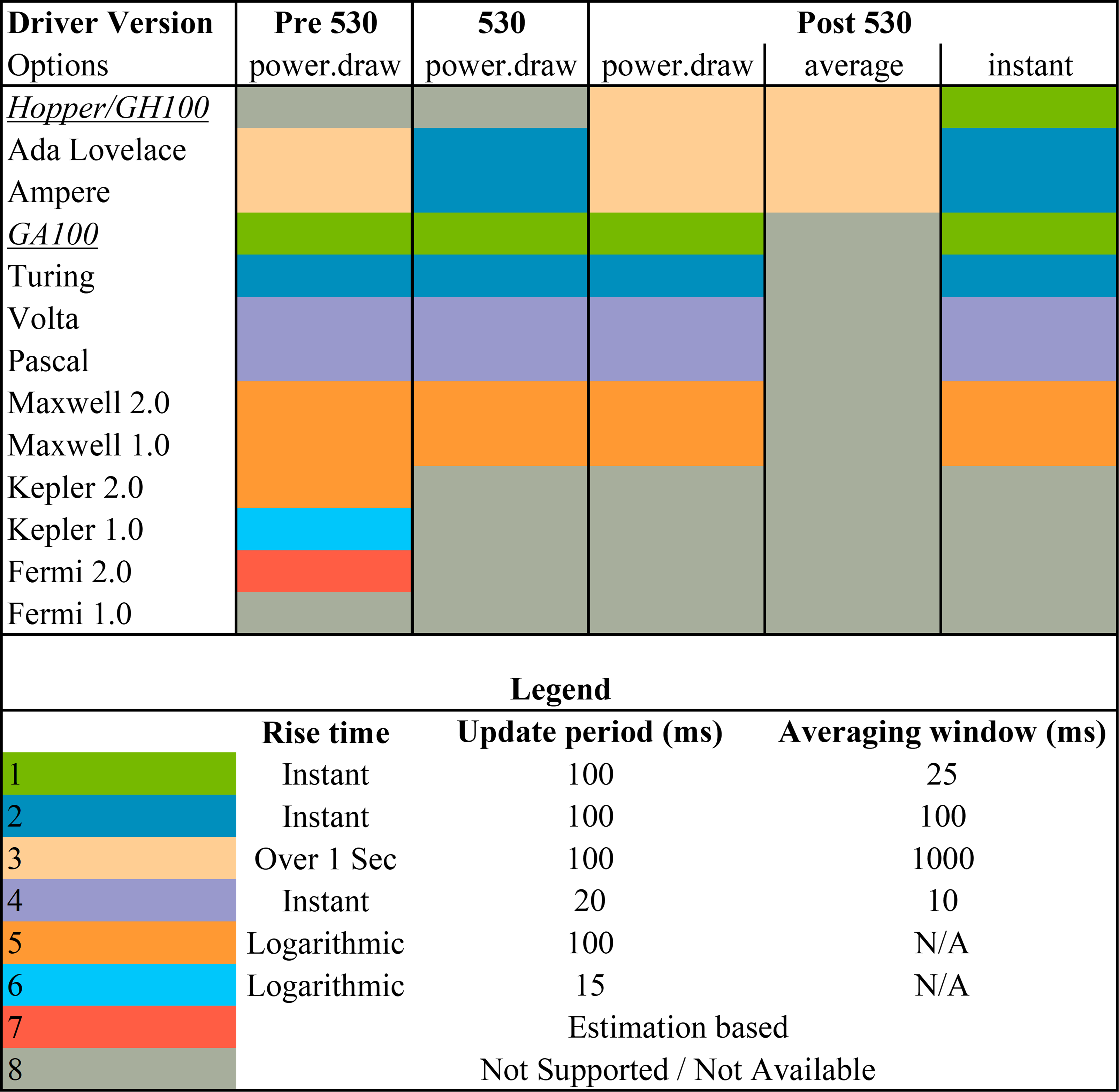}
    \caption{Results summarised from all tested GPUs with different driver versions. The `Fermi' generation GPUs either didn't support power measurement or was estimation based. `Kepler' and `Maxwell' generation GPUs show a logarithmic growth like transient response, which the power update frequency are different. `Volta' and `Pascal' GPUs all have instant rise time, 20ms power update period and 10ms averaging window. `Turing' GPUs have instant rise time, 100ms update period and 100ms averaging window. The situation becomes more complicated for `Ampere' and newer generations. A100 cards have a 25ms averaging window for all driver versions. For the rest of other `Ampere' GPUs and the `Ada Lovelace' GPUs, on drivers released before March 2023, the averaging window is the past 1 second. On driver version 530 the averaging window changed to 100ms, and after that its switch back to 1 second, whereas the 100ms version is the newly added instant power draw option. Lastly, for the H100, the instant option gives a averaging window of 25ms, while the average and normal option gives an average over 1 second.}
    \label{overall}
\end{figure}

Overall, the error exists in 2 domains, power and time. Error in the power domain is predominantly due to the intrinsic tolerances of shunt resistors. Error in the time domain comes from the strange behaviours of nvidia-smi's internal data processing. For the two different measurement perspective outlined in the background section:
\begin{itemize}[leftmargin=*]
    \item Power: The error in power measurements are proportional (roughly $\pm$5\%) rather than a flat value ($\pm$5W claimed by NVIDIA). On modern GPUs capable to drawing 700W this could lead to a $\pm$30W of over/underestimation. Comparing the power consumption of different tasks on the same GPUs is still reliable thanks to the near perfect linear relationship. However, since the error margin is random, comparing results from different GPUs would be erroneous. The 30W of additional error may seem insignificant, but can accumulated in computing systems with thousands of GPUs, although statistically the error could (but not guaranteed to) average out for large quantities.
    \item Energy: Energy is the product of power and time. In addition to the error in power, it is also suspect to the issues in time domain caused by nvidia-smi's internal data processing. These issues in the time domain is much more convoluted, and we tried to analyze and find solutions to correct these issues in the next chapter.
\end{itemize}

\section{Energy Measurement Evaluation}
In this section, we explore good practice for energy measurement, and verify our results using real world workloads.

\subsection{Exploration}
Common practice when making measurements is to conduct multiple repetitions of the measurement and take the average. Undertaking too few repetitions might lead to larger error, while too many repetitions can result in wasted time and energy. Our experiment aims to determine the necessary number of repetitions for accurate energy consumption measurements of a program, balancing accuracy, precision, and efficiency. We also intend to apply insights about nvidia-smi behaviour from the previous section to enhance measurement accuracy.

We explore three distinct cases: when the averaging period is the same, longer or shorter than the power update frequency. The first two cases are examined on the RTX 3090 with the instant and average power draw options. The third case is tested on the A100. For each case we test a short, medium and long benchmark load periods, corresponding to 25\%, 100\% and 800\% of the power update frequency. The reason for continuing use of the benchmark load is because of its controllable nature and the challenges its repeated high/low power states pose to nvidia-smi's measurement, thereby representing a worst-case scenario. We did not test GPUs with a logarithmic growth transient response as they've already been studied by others, and these GPUs are nearing, if not already, end of life.

\paragraph{Case 1} Our experiment tests a varying number of repetitions for each workload, conducting 32 trials for each. We introduced a random 0-1 second delay between trials. Figure~\ref{100,100} plots the mean percentage error and standard deviation of each trial's error against the ground truth. The dotted line and cross illustrate the initial nvidia-smi results, from integrating power over the kernel execution period. Fewer repetitions significantly underestimated the power consumption.

\begin{figure}[h!]
    \centering
    \includegraphics[width=0.75\columnwidth]{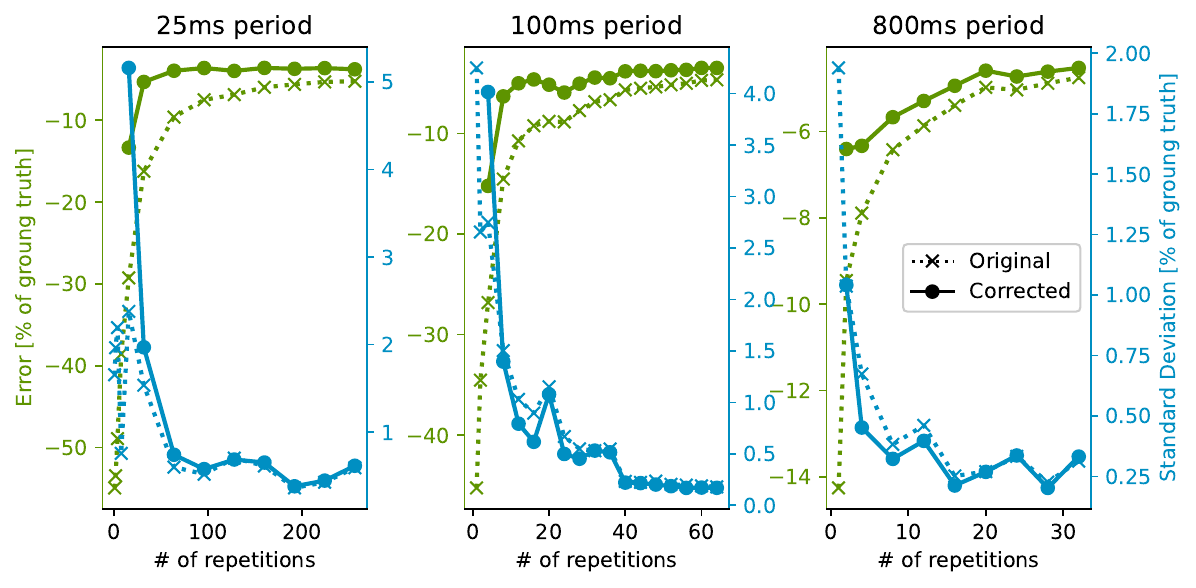}
    \caption{Case one: averaging window same as power update period. Results are tested on a RTX 3090 with the power.draw.instant option. In this case the averaging window and power update period are both 100ms.}
    \label{100,100}
    \vspace{-0.4cm}
\end{figure}

As repetition counts increased, the mean error decreased, stabilising at around -5\%, indicative of the GPU's steady-state power draw error margin. The standard deviation also diminished, converging to about 0.5\%. This suggests that a higher number of repetitions enhances measurement accuracy to a level comparable with the inherent accuracy of the power measurement hardware, while also improving precision. However, excessive repetitions do not further benefit accuracy or precision and lead to unnecessary time and energy expenditure.

The solid line and dots show the energy measurements corrected using insights from the previous section. Considering this GPU's 250-millisecond rise time, we disregarded the initial 10 repetitions covering this time frame. In addition, the 100ms boxcar averaging window and power update period suggests that the reported power draw actually corresponds to the GPU activity from 100ms prior. Thus the power draw from nvidia-smi is shifted 100ms earlier. This correction showed that accurate energy measurements can be achieved with fewer repetitions and attain higher accuracy levels more rapidly.

\paragraph{Case 2} The default setting for reporting average power over the previous second on 'Ampere', 'Ada', and 'Hopper' GPUs makes this scenario highly relevant for most users in upcoming years. Fig.~\ref{1s,100} shows the results obtained from experiments conducted for this case. A similar trend was observed, where the accuracy and precision increased as the number of repetitions increased. However, due to the gradual ramp-up of power consumption over one second, it required a greater number of repetitions for the accuracy to increase and converge. Yet, by discarding the initial 1250ms (250ms rise time + 1s average time) of data, we attained accuracy levels comparable to Case 1 with even fewer repetitions.

\begin{figure}[h!]
    \centering
    \includegraphics[width=0.75\columnwidth]{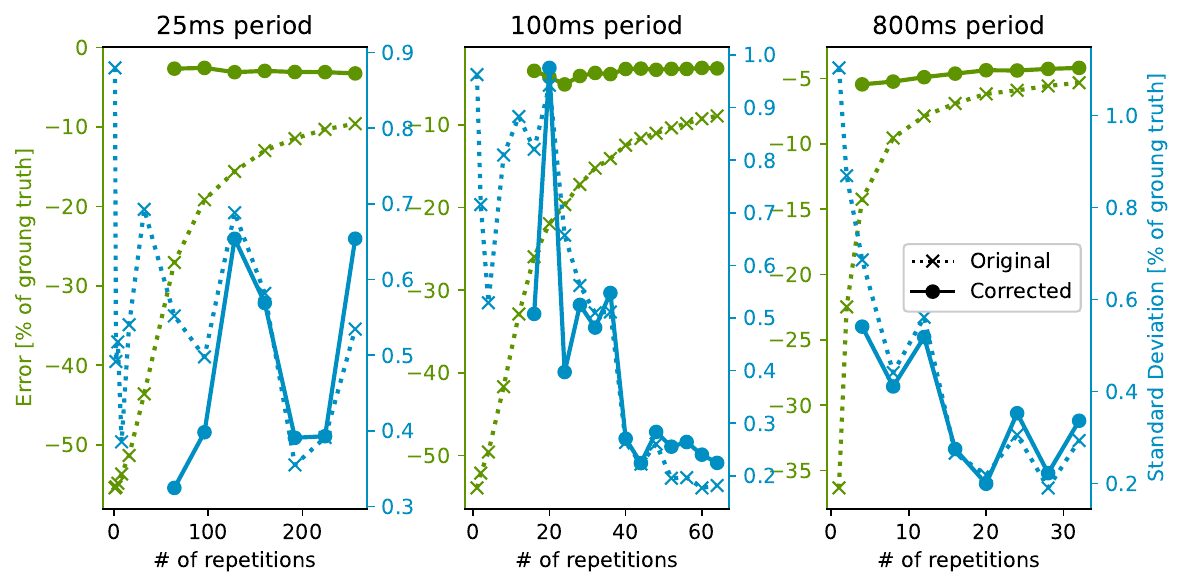}
    \caption{Case Two: averaging window is longer than the power update period. Results are tested on a RTX 3090 with the power.draw.instant option. In this case the averaging window is 1000ms and the power update period is 100ms.}
    \label{1s,100}
    \vspace{-0.45cm}
\end{figure}

\paragraph{Case 3} On H100 and A100, nvidia-smi only reports the average of the past 25ms every 100ms, hence 75\% of the information is not being captured. Similarly, on `Volta' and `Pascal' cards 50\% of information is lost due to the 10/20ms reporting interval. To overcome this issue, we propose a measurement strategy incorporating controlled delays between repetitions to shift the 'phase' of GPU activity, thereby exposing different workload segments to the measurement window.

Fig.~\ref{25,100} presents results obtained from testing on the A100 GPU. For a given repetition, such as 64, the 0 shifts test means 64 consecutive repetitions, just as for Case 1 and 2. A 4 shifts test indicates a 25ms delay inserted after every 16 consecutive repetitions, while 8 shifts meant a 25ms delay following every 8 repetitions. The correction applied here is discarding the first 100ms of repetitions according to the rise time of the GPU.

In the 25ms period benchmark load, the entire activity aligned with the 25ms averaging window, resulting in accuracy and precision trends similar to the previous cases. However, for the 100ms and 800ms period loads, only a quarter of the activity was captured. This, combined with nvidia-smi's unpredictable start time, led to significant random deviations in measured versus actual power consumption. For the 100ms load, the measurement error's standard deviation reached up to 30\%. Implementing controlled 25ms delays, either 4 or 8 times, significantly reduced this standard deviation to below 5\% and stabilised the mean percentage error, thus improving both the accuracy and precision of our measurements.

\begin{figure}[h!]
    \centering
    \includegraphics[width=0.6\columnwidth]{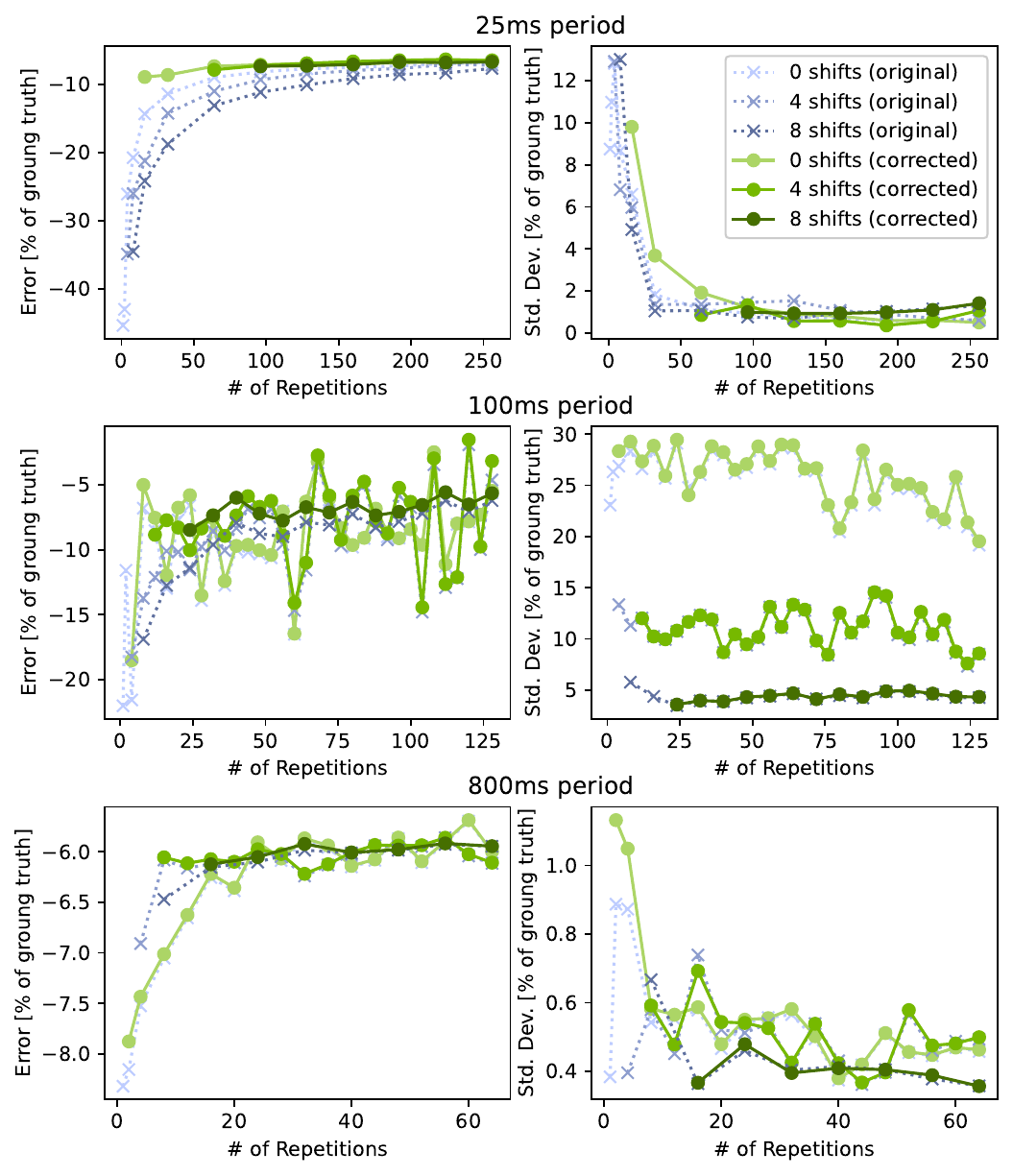}
    \caption{Case Three: averaging window is longer than the power update period. Results are tested on a A100 with the power.draw.instant option. In this case the averaging window is 25ms out of the 100ms power update period.}
    \label{25,100}
    \vspace{-0.45cm}
\end{figure}

Upon analysing results from three distinct scenarios, we propose the following measurement good practice:
\begin{enumerate}[leftmargin=*]
    \item Execute the target program for 32 consecutive iterations or until a minimum runtime of 5 seconds is reached. If data loss occurs due to a small averaging window, insert 8 controlled delays evenly spaced within the repetitions.
    \item Perform four separate trials, introducing a randomised delay between each.
    \item Post processing the collected data by discarding repetitions occurred during rise time, and shift the data to synchronise with the activity of the GPU.
\end{enumerate}

\subsection{Benchmark Selection}
We aimed to examine the accuracy of using nvidia-smi to measure the energy consumption in real-world workloads. We selected nine benchmarks representing a variety of applications, detailed in Table.~\ref{table:benchmarks}. These were configured to cover a range of execution times and energy consumption levels. However, benchmarks requiring graphics capabilities, like SPECViewPerf, were omitted because many of the data center GPUs we tested lack these features.

\vspace{-0.2cm}
\begin{table}[htbp!]
\caption{Selected Benchmarks}
\centering
\label{table:benchmarks}
\begin{tabular}{l c c c}
\toprule
\textbf{Source} & \textbf{Benchmark} & \textbf{Application} \\ 
\midrule
NV Library & CUBLAS & Linear Algebra \\
 & CUFFT &  Signal Processing\\
 & nvJPEG & Image Compression \\
\midrule
Domain Specific & Stereo Disparity & Computer Vision \\
 & Black-Scholes & Computational Finance \\
 & Quasi-random Gen &  Monte Carlo, etc\\
\midrule
MLPerf & ResNet-50 & Image Classification \\
 & RetinaNet & Object detection \\
 & Bert & Natural Language Processing \\
\bottomrule
\end{tabular}
\end{table}

\subsection{Energy Measurement Results}
\begin{figure}[h!]
    \centering
    \includegraphics[width=0.75\columnwidth]{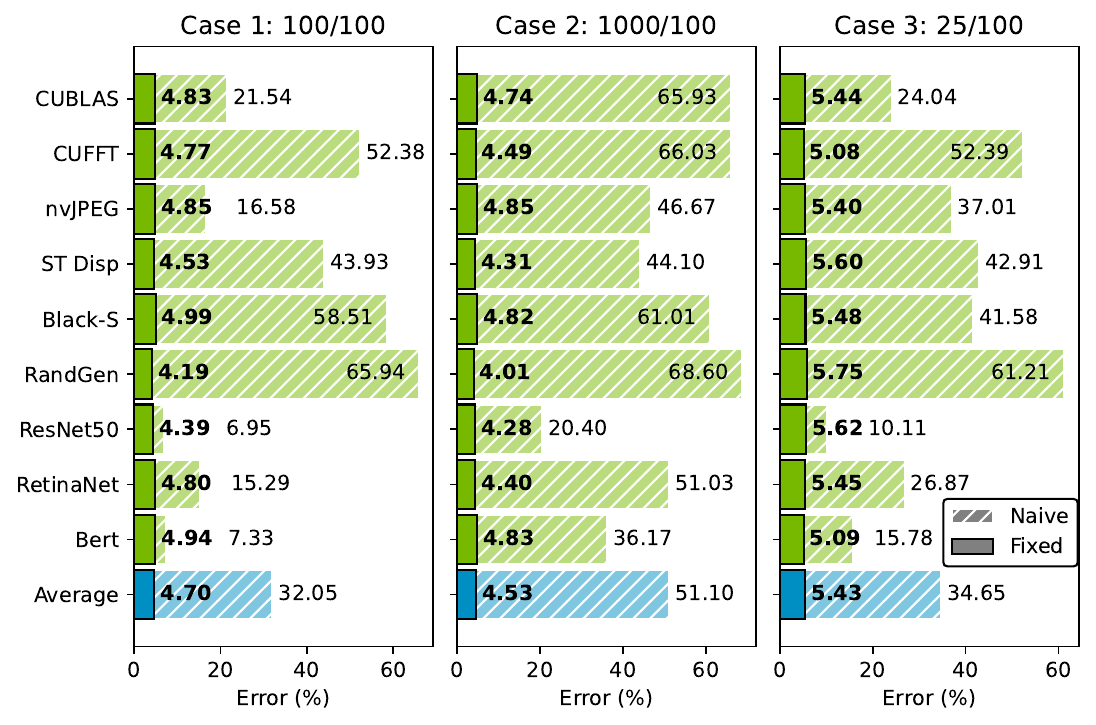}
    \vspace{-0.3cm}
    \caption{The error in energy of the 9 selected benchmarks measured from naive approach and good practice compared to using the PMD for the 3 different cases: averaging window is the same as (100/100), longer than (1000/100), or shorter than (25/100) the power update period. On average, following the good practice can reduce the measurement error by 34.38\%, from 39.27\% to 4.89\%. The standard deviation of the reduced error for each case are 0.25\%, 0.28\% and 0.21\%}
    \label{energy_meas}
\end{figure}

We measured the energy consumption of nine different workloads using a naive method of just running the workload once and take nvidia-smi's reported numbers as granted, and the above-proposed measurement practice for each case. The results were compared with energy calculated using PMD data, and the errors are depicted in Fig.~\ref{energy_meas}. Error in the energy measurement from the naive method highly varies up to 70\% randomly among the different workloads and cases. Nonetheless, by following the good practice, the error can be reduced to around 5\% across the board. Moreover, the standard deviation of the good practice error across all cases and workloads is 0.25\%, indicating that the proposed measurement practice is stable across different GPUs and workloads.

Furthermore, these error margins align with the power measurement error of each GPU. For cases 1 and 2, conducted on the RTX 3090, the errors were -4.70\% and -4.53\% respectively. For case 3, conducted on the A100, the error was -5.43\%. Applying the power measurement error gradient and offset as a transform on the nvidia-smi data will reduce the error to nearly zero. This indicates that we have corrected all the errors in the time domain caused by the behaviours of nvidia-smi. Despite this, it is impossible to mitigate the error caused by the random component resistance tolerance of a physical shunt resistor on the GPU's PCB.

\section{The Grace Hopper Superchip Evaluation}
The GH200 Grace Hopper Superchip, NVIDIA's latest flagship, combining a powerful Hopper GPU and a 72-core Arm based Grace CPU in the same package, leveraging a high-bandwidth NVLink interconnect~\cite{GraceHopperSuperchipDataSheet}. A key highlight is its power efficiency, attributed to the Arm CPU and LPDDR5X memory. The University of Bristol has secured £220 million to acquire 5,448 GH200 chips~\cite{bristol_GH200}, and Germany is set to purchase 24,000 GH200 for the JUPITER supercomputer~\cite{germanyGH200}. Given its claimed power efficiency and global demand, it's necessary to assess GH200's power measurement capabilities.

\begin{figure}[h!]
    \centering
    \includegraphics[width=0.7\columnwidth]{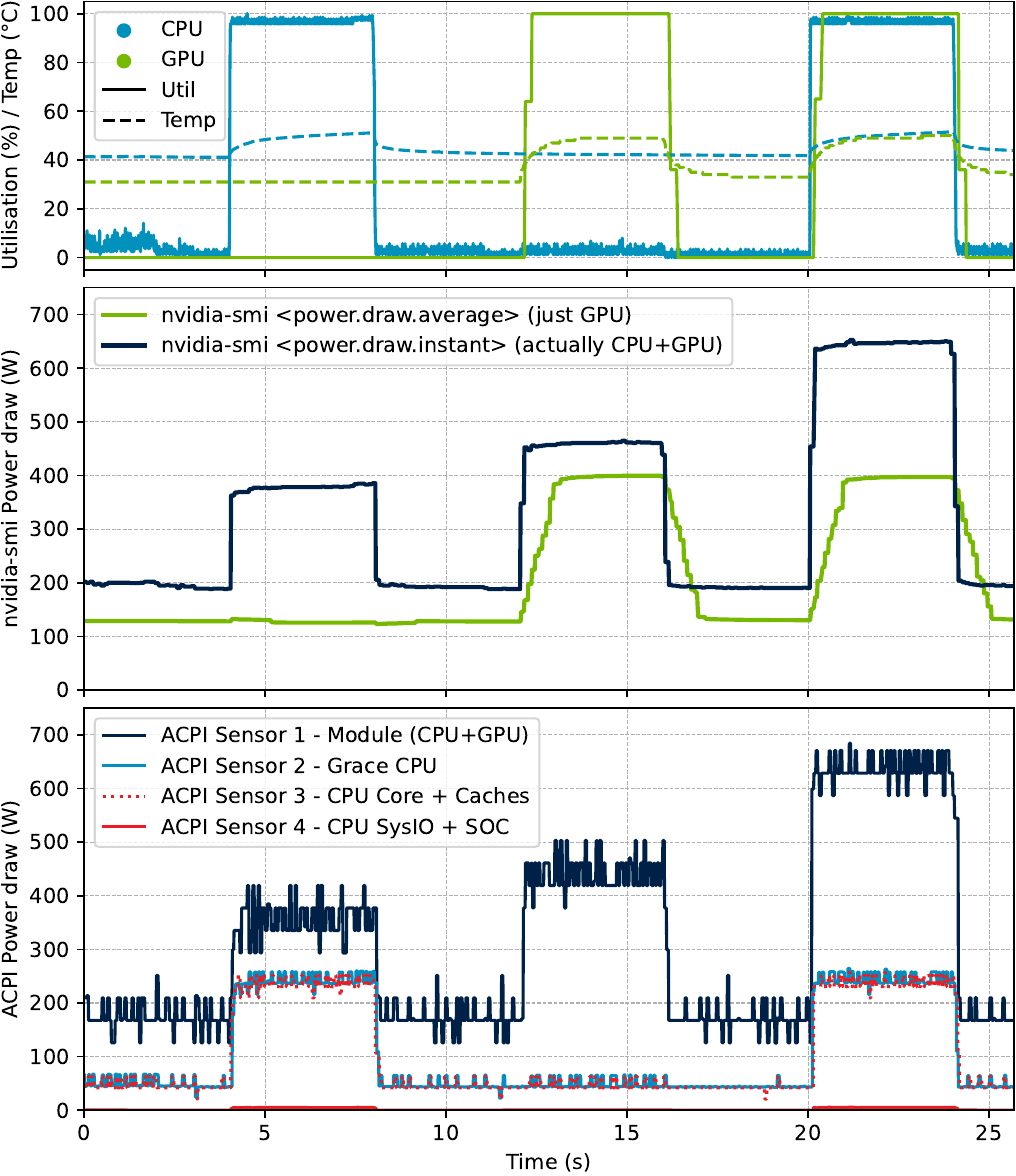}
    \caption{Experiment of running a CPU and a GPU separately and then simultaneously. TOP: the utilisation and temperature of the CPU and GPU, indicating when the CPU, GPU and both are running. MIDDLE: Power draw output of nvidia-smi. The \textless{}Average\textgreater{} power is doing what it should do, but the \textless{}Instant\textgreater{} power reacts to both CPU and GPU activity, indicating its a measurement of the whole GH200. BOTTOM: Power draw data of ACPI. The overall shape seems to follow nvidia-smi, however the waveform is overall extremely flat while having seemingly discrete fluctuations over 100W.}
    \label{GH200} 
\end{figure}

First of all, we found  discrepancies between the `Average' and `Instant' power query options in nvidia-smi, with the `Instant' readings consistently exceeding `Average' under various conditions. Typically, on other GPUs, these metrics should align during idle periods or under sustained loads, as both aim to measure GPU power—`Average' being a one-second running average of `Instant'. Suspecting that `Instant' might encompass additional system components, we conducted an experiment involving separate and simultaneous CPU and GPU loads, as depicted in Figure~\ref{GH200}. The findings indicate that the `Instant' measurement, ostensibly representing only GPU power consumption, actually reflects the combined power draw of the GPU, CPU, and presumably the DRAM as well.

After evaluating both the GPU and CPU power draw using our benchmark suite, it was observed that both readings update every 100ms. However, the measurement window encompasses only the last 20ms for GPU activity and 10ms for CPU activity. Consequently, 80\% of GPU activity and 90\% of CPU activity are overlooked. This is arguably another step backwards from the already substantial 75\% oversight in power activity measurement found in the A100 and H100 GPUs. Without the insights provided by our benchmark suite, one might assume continuous power monitoring, whereas, in reality, only a fraction (10\%) of the activity is being measured.

As a part of the Grace CPU, NVIDIA has incorporated power sensors accessible via the ACPI interface to record the average power every 50ms across different domains of the GH200~\cite{GraceHopperPerformanceTuning2024}. While performing the CPU/GPU load testing, we also collected power data using this method, as illustrated in Fig.~\ref{GH200}. The results generally align with those obtained from nvidia-smi, albeit with an anomalously flat power draw profile punctuated by discrete 'noise' fluctuations, exhibiting swings exceeding 100W. Without additional information, it remains challenging to definitively ascertain the methodology behind this power measurement approach.

Overall, the power measurement capabilities of the Grace Hopper Superchip leave much to be desired. Issues range from misleading power query options to even less reliable `part-time' power measurement compared to its predecessors, the A100 and H100 GPUs. Additionally, the ACPI power draw data exhibits 'noise' with fluctuations exceeding 100W, further complicating accurate power consumption analysis.

\section{Related Work}
Burtscher et al.~\cite{burtscher2014measuring} were among the first to investigate the GPU on-board power sensors. They tested 2 n-body algorithms on 3 Tesla K20c/m/x GPUs, and gathered the power readings using the NVML Library. They discovered that the sampling period is 15ms, and the power draw is distorted. We made the same observation when testing the K40 GPU. They modelled the distortion using the capacitor charging formula, and proposed a methodology to correct the distortion. Aslan et al.~\cite{aslan2022study} performed a similar experiment on 2 embedded GPUs: NVIDIA Jetson TX2 (Pascal) and AGX Xavier (Volta), and observed the same ``capacitor charging'' effect.

Fahad et al.~\cite{fahad2019comparative} compared the on-board power sensors against a WattsUp external power meter. The WattsUp meter gose between the mains plug and socket, and measures the power for the entire computer. They tested a GEMM MatMul and a 2D FFT algorithm on a K40 and a P100 PCIe GPU. They found that the error of the energy profile obtained by querying NVML could be as high as 73\% of the external power meter.

Sen et al.~\cite{sen2018quality} also tested the Tesla K20c GPU from a different perspective. The generated a similar square-wave like load using a matrix multiply kernel, and gather power data from NVML and a PowerInsight external power meter~\cite{laros2013powerinsight} capable of capturing the GPU power consumption at 3KHz sampling frequency. They found that the external power meter has a higher quality compared to NVML as it better reproduced the square-wave shape. They also observed the `capacitor charging' effect, although they modelled it using a moving average filter. Jay et al.~\cite{jay2023experimental} also compared a range of software tools that helps the power measurement by interface with NVML internally, including CarbonTracker~\cite{henderson2020towards}. They tested several benchmarks on a DGX server with 8 V100s, and compared the results against a Omegawatt external power meter~\cite{omegawatt} sampling at 1Hz. The focus of this work is qualitatively comparing the functionalities the different software tools.

The GPUs analysed in previous studies were limited and outdated, often including only a few models like the six-year-old V100 and the eleven-year-old K20. In contrast, our research encompasses over 70 GPUs across all product lines, form factors, and 12 architectural generations, with a focus on recent models such as the H100, A100, RTX 4090, and 3090.

Previous works generally assessed apparent measurement results superficially and did not delve into the underlying mechanisms, leading to largely speculative conclusions with limited practical utility. Our study comprehensively reverse-engineered the entire measurement workflow, from electronic circuits to data post-processing, using various experiments and analyses. This enabled us to accurately assess measurement accuracy across different scenarios and offer concrete guidelines for effective use of nvidia-smi in power measurement.

\section{Conclusion}
In this paper, we investigated the hardware and software mechanisms of NVIDIA GPU's on-board power sensor. We designed a suite of micro-benchmarks to characterise the behaviour of nvidia-smi, and tested on a exhaustive set of over 70 GPUs. We further examined the power and energy measurement accuracy of the built-in sensor in comparison to the external power meter. Subsequently, we proposed a measurement standard to ensure accurate and precise measurements using nvidia-smi. The micro-benchmark is available on GitHub\footnote{\url{https://github.com/JimZeyuYang/GPU_Power_Benchmark}}, alongside a spreadsheet detailing our GPU test results, which will be updated with future GPU releases. We hope that a robust understanding of the NVIDIA GPU's on-board power sensor's mechanisms and accuracy will empower researchers to use it confidently for accurate power measurements, promoting research in energy efficient computing on GPUs.

\section*{Acknowledgments}
The authors would like to acknowledge the use of the University of Oxford Advanced Research Computing (ARC) facility in carrying out this work. http://dx.doi.org/10.5281/zenodo.22558

\bibliographystyle{unsrt}  
\bibliography{refs}

\end{document}